 \documentclass[final,5p,times,twocolumn,authoryear]{elsarticle}

\usepackage{amssymb}
\usepackage{lipsum}
\usepackage{amsmath}
\usepackage{mathrsfs}

\usepackage[colorlinks=true, allcolors=blue]{hyperref}
\usepackage{orcidlink}
\journal{Physics Letters B}

\begin{document}

\begin{frontmatter}


\title{Conventional vs. modified GTD metrics: Survival of modified GTD metrics in AdS spacetime and thermodynamic ensembles}

\author[first,second]{Gunindra Krishna Mahanta \, \orcidlink{0000-0003-4544-2585}}
\ead{guninmohantaba@gmail.com}

\affiliation[first]{organization={Astrophysical Sciences Division},%Department and Organization
            addressline={Bhabha Atomic Research Centre}, 
            city={Mumbai},
            postcode={400085}, 
            state={Maharashtra},
            country={India}}

\affiliation[second]{organization={Homi Bhabha National Institute},%Department and Organization
            addressline={Anushaktinagar}, 
            city={Mumbai},
            postcode={400094}, 
            state={Maharashtra},
            country={India}}

\begin{abstract}
Thermodynamic geometry provides a powerful framework for probing the microscopic structure of thermodynamic systems. Among its formulations, Geometrothermodynamics (GTD) has been widely applied to black hole thermodynamics, owing to its Legendre-invariant construction. However, recent work \citep{Mahanta_2025} has shown that conventional GTD metrics fail to encode essential physical boundaries of thermodynamic phase space. By modifying the conventional metric structure, three new GTD metrics were previously introduced, which successfully capture these boundaries in regular spacetime. Whether such modified metrics remain viable in different spacetime backgrounds and under changes of thermodynamic ensemble has remained an open question. In this work, I address this issue by investigating the behavior of modified GTD metrics in AdS spacetime and across different thermodynamic ensembles in the framework of Bardeen AdS black hole. An analysis of thermodynamic geodesics demonstrates that the modified GTD metrics consistently respect physical boundaries of the phase space, in contrast to conventional GTD metrics. This behavior is preserved in AdS spacetime and under Legendre transformations, establishing the robustness and universality of the modified GTD framework.

\end{abstract}


\end{frontmatter}

%\tableofcontents


\section{\label{sec:intro}Introduction}

Thermodynamic geometry has emerged as a powerful framework for probing the microscopic structure of black holes (BHs). While the macroscopic thermodynamic properties of a BH can be derived from its spacetime geometry using the four laws of black hole thermodynamics \citep{Bardeen_1973_4_law}—an insight deeply rooted in the geometric formulation of gravity \citep{Padmanabhan_2002,Jacobson_1995,Wald_1993}—the inverse problem of uncovering the microscopic interactions among the underlying degrees of freedom requires a different geometric approach. Thermodynamic geometry provides such a framework by encoding thermodynamic information into a geometric structure defined on the thermodynamic phase space.

The central idea of thermodynamic geometry is to introduce a metric on the space of equilibrium thermodynamic states and to extract microscopic information through geometric quantities such as curvature and geodesics, using tools analogous to those of general relativity. Early conceptual foundations were laid by Gibbs \citep{Andrews_1929} and Carath\'eodory \citep{Caratheodory_1909}, with further geometric developments by Hermann \citep{Hermann_1973} and Mrugala \citep{Murgala_1978,Murgala_1985}. A concrete realization of this idea was provided by Weinhold, who defined a thermodynamic metric as the Hessian of the internal energy \citep{Weinhold_1975}. Subsequently, Ruppeiner introduced an entropy-based metric motivated by thermodynamic fluctuation theory \citep{Ruppeiner_1979}. These approaches have been extensively applied to a wide class of thermodynamic systems, including black holes, to study phase transitions, thermodynamic length, curvature, and geodesic structure \citep{Salamon_1980,Salamon_1984,Salamon_1985,Gilmore_1981,Feldmann_1985,Kumar_2012}.

Despite their success, the Weinhold and Ruppeiner formalisms suffer from a fundamental limitation: they are not invariant under Legendre transformations. Since thermodynamic behavior depends on the choice of ensemble, this lack of Legendre invariance can lead to inconsistent or even incorrect physical predictions. A well-known example is the absence of a phase transition signal in the Weinhold geometry of the Kerr–AdS black hole, in contradiction with established results in black hole thermodynamics \citep{Aman_2003}. This inconsistency highlights the necessity of a thermodynamic metric framework that is invariant under changes of thermodynamic potential.

To address this issue, Quevedo introduced the Theory of Geometrothermodynamics (GTD) \citep{Quevedo_2007}, which provides a Legendre-invariant geometric formulation of thermodynamics. The main achievement of GTD theory is that this theory introduces a Legendre invariant set of metric which allows to define a compatible metric structure in a thermodynamics phase space and thermodynamics equilibrium space consistently. Owing to its Legendre-invariant nature, GTD has been successfully applied to a variety of thermodynamic systems, including the van der Waals gas \citep{Quevedo_2007,Quevedo_2022} and black hole thermodynamics \citep{Alvarez_2008,Quevedo_2008a,Quevedo_2008b,Vazquez_2010,Gogoi_2023}.

However, despite these successes, the conventional GTD metrics possess intrinsic limitations. In my earlier work \cite{Mahanta_2025} (hereafter Paper I), I demonstrated that thermodynamic geodesics defined by conventional GTD metrics fail to recognize essential physical boundaries of a thermodynamic system, such as the temperature-vanishing curve (the curve which separate positive temperature region from negative temperature region) and the spinodal curve (which separate positive specific heat region from negative specific heat region). This failure indicates that the conventional GTD metric structure does not fully encode the physical constraints of black hole thermodynamics. Motivated by this observation, I introduced modified GTD metrics in Paper I and showed, in the context of the Bardeen regular black hole, that these modified metrics successfully capture all physical boundaries of the thermodynamic phase space. Based on the behavior of thermodynamic geodesics and curvature scalars, I argued that the modified GTD metrics provide a more faithful geometric description of black hole thermodynamics.

While the modified GTD metrics were shown to be effective in regular spacetimes and within the canonical ensemble, their general validity remained an open question. In particular, it is not a priori clear whether these modified metric structures survive in asymptotically AdS spacetimes or retain their predictive power under changes of thermodynamic ensemble.

The present work addresses these issues by focusing on two central questions:
\begin{enumerate}
    \item Do the modified GTD metrics remain well behaved in AdS spacetime?
    \item Are the modified GTD metrics capable of consistently identifying physical boundaries under Legendre transformations, i.e., in different thermodynamic ensembles?
\end{enumerate}

To address the first question, I investigate the behavior of thermodynamics geodesic in the thermodynamics spacetime in the framework of Bardeen AdS BH in canonical ensemble, to address the second, I perform a Legendre transformation of the thermodynamic potential and analyze the system in the grand canonical ensemble. This combined analysis allows for a stringent test of the robustness and universality of the modified GTD framework.

In the semiclassical treatment of AdS black hole thermodynamics, the partition function is defined through the Euclidean path integral, 
\begin{equation}
Z \sim e^{-I_E [g_{cl}]} \nonumber
\end{equation}
from which the thermodynamic potentials are obtained (where $g_{cl}$ is the classical black hole solution, and $I_E [g_{cl}]$ is the Euclidean action evaluated on that solution) \citep{Gibbons_1977,Hawking_1983}. The stability conditions in a given ensemble are determined by the convexity properties of the corresponding thermodynamic potential, which in turn reflect the behavior of the partition function. The GTD construction adopted here does not require explicit knowledge of the partition function; rather, it encodes these stability properties geometrically through the Legendre-invariant metric structure. In this sense, the geometric analysis is consistent with the underlying ensemble-dependent partition function framework.\\

The paper is organized as follows. In Sec.~\ref{sec:Bardeen_Ads}, I briefly review the thermodynamics of the Bardeen AdS black hole. Section~\ref{sec:con_vs_mod} introduces the conventional and modified GTD metric structures. A brief discussion of thermodynamic geodesics is presented in Sec.~\ref{sec:geo_eq}. The behavior of thermodynamic geodesics in the canonical ensemble is analyzed in Sec.~\ref{sec:geo_in_can}, while Sec.~\ref{sec:GC_ens} discusses the thermodynamics in the grand canonical ensemble. The corresponding geodesic analysis is presented in Sec.~\ref{sec:geo_in_GC}. The results are summarized and discussed in Sec.~\ref{sec:Result}.

\section{\label{sec:Bardeen_Ads}Bardeen A\MakeLowercase{d}S black hole}
Under the assumption of spherical symmetry, the line element of the Bardeen AdS black hole can be written as \citep{Fernando_2017,Tzikas_2019}
\begin{equation}
ds^{2}=-f(r)\,dt^{2}+\frac{1}{f(r)}\,dr^{2}+r^{2}d\Omega ,
\end{equation}
where
\begin{equation}
f(r)=1-\frac{2mr^{2}}{(g^{2}+r^{2})^{3/2}}+\frac{r^{2}}{l^{2}} .
\end{equation}
Here $g$ denotes the magnetic charge and $l$ is the AdS radius. The event horizon radius $r_{h}$ is determined by the condition $f(r_{h})=0$, which yields
\begin{equation}
m=\frac{(g^{2}+r_{h}^{2})^{3/2}}{2r_{h}^{2}}\left(1+\frac{r_{h}^{2}}{l^{2}}\right).
\label{eq:m_rh}
\end{equation}

Using the Bekenstein–Hawking area law \citep{Bekenstein_1973}, $S=\pi r_{h}^{2}$, I introduce the rescaled entropy variable $s=S/\pi=r_{h}^{2}$. Equation~(\ref{eq:m_rh}) can then be expressed as
\begin{equation}
m=\frac{(s+g^{2})^{3/2}}{2s}\left(1+\frac{s}{l^{2}}\right).
\label{eq:m_s}
\end{equation}

It is convenient to work with dimensionless thermodynamic variables, obtained by the scalings $m\rightarrow m/l$, $s\rightarrow s/l^{2}$, and $g\rightarrow g/l$. With these rescalings, Eq.~(\ref{eq:m_s}) reduces to
\begin{equation}
m=\frac{1}{2s}(s+g^{2})^{3/2}(1+s),
\label{eq:fundamental_eq}
\end{equation}
which serves as the fundamental equation of the Bardeen AdS black hole. This expression shows that the system constitutes a two-dimensional thermodynamic manifold with extensive variables $(s,g)$.

The corresponding intensive variables are obtained from the fundamental equation as
\begin{eqnarray}
t=\left(\frac{\partial m}{\partial s}\right)_{g}
=\frac{\sqrt{s+g^{2}}\left(-2g^{2}+3s^{2}+s\right)}{4s^{2}}, \\
\phi=\left(\frac{\partial m}{\partial g}\right)_{s}
=\frac{3g(s+1)\sqrt{s+g^{2}}}{2s}.
\label{eq:phi}
\end{eqnarray}

The specific heat at constant magnetic charge is given by
\begin{equation}
c_{g}=t\left(\frac{\partial s}{\partial t}\right)_{g}
=\frac{\left(\partial m/\partial s\right)_{g}}
{\left(\partial^{2}m/\partial s^{2}\right)_{g}}
=\frac{2s(s+g^{2})(-2g^{2}+3s^{2}+s)}
{8g^{4}+4g^{2}s+s^{2}(3s-1)} .
\end{equation}

In the canonical ensemble, the physical thermodynamic region of the black hole is bounded by two characteristic curves: the spinodal curve and the temperature-vanishing curve. The spinodal curve separates regions of positive and negative specific heat and corresponds to a second-order phase transition. Equivalently, it can be identified as the locus of Davies points at which the specific heat diverges \citep{Davies_1978}, obtained by the condition $c_{g}\to\infty$. The temperature-vanishing curve separates positive- and negative-temperature regions and is determined by $t=0$.

The explicit forms of these curves are
\begin{eqnarray}
g=\frac{\sqrt{s}\sqrt{3s+1}}{\sqrt{2}},
\label{eq:t=0}\\
g=\sqrt{\frac{1}{4}\sqrt{3}\sqrt{s^{2}-2s^{3}}-\frac{s}{4}} .
\label{eq:spinodal}
\end{eqnarray}

\begin{figure} \centering \includegraphics[width=\linewidth]{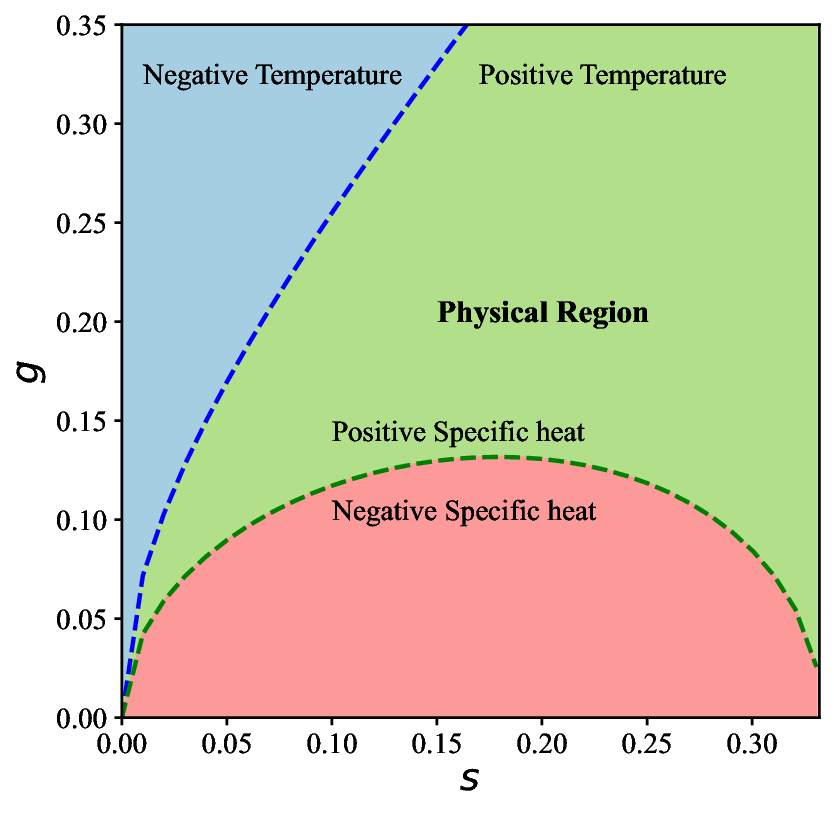} \caption{$s$--$g$ thermodynamic phase space of the Bardeen AdS black hole. The green dotted curve denotes the spinodal line separating regions of positive and negative specific heat, while the blue dotted curve represents the temperature-vanishing line. The green shaded area corresponds to the physical region with positive temperature and positive specific heat; red and blue shaded regions indicate negative specific heat and negative temperature, respectively.} \label{fig:physical_region_canonical} \end{figure}

The physical region of the Bardeen AdS black hole in the canonical ensemble is therefore defined by the domain in which both the temperature and the specific heat are positive, as illustrated in Fig.~\ref{fig:physical_region_canonical}. 

In the following sections, I analyze the behavior of thermodynamic geodesics defined by both conventional and modified GTD metrics and provide a qualitative comparison between the two frameworks.

\section{\label{sec:con_vs_mod}Conventional and Modified GTD metrics}

Conventional GTD metrics are given by \citep{Quevedo_2019_QBH, Quevedo_2024}
\begin{align}
\mathcal{G}^I &= \sum_{a,b,c=1}^{n} \left( \beta_c E^c \frac{\partial \Phi}{\partial E^c} \right) \frac{\partial^2 \Phi}{\partial E^a \partial E^b} \, dE^a \, dE^b \label{eq:gI}\\
\mathcal{G}^{II} &= \sum_{a,b,c,d=1}^{n} \left( \beta_c E^c \frac{\partial \Phi}{\partial E^c} \right) \eta^d_a \frac{\partial^2 \Phi}{\partial E^b \partial E^d} \, dE^a \, dE^b \label{eq:gII}\\
\mathcal{G}^{III} &= \sum_{a,b=1}^{n} \left( \beta_a E^a \frac{\partial \Phi}{\partial E^a} \right) \frac{\partial^2 \Phi}{\partial E^a \partial E^b} \, dE^a \, dE^b \label{eq:gIII}
\end{align}

While, the modification introduced to these conventional metrics in Paper I is given as

\begin{align}
\mathcal{G}^I_{\mathrm{mod}} &= \sum_{a,b,c=1}^{n} \left( \beta_c E^c \frac{\partial \Phi}{\partial E^c} \delta^c_1  \right) \delta^b_a \frac{\partial^2 \Phi}{\partial E^a \partial E^b} \, dE^a \, dE^b 
    \label{eq:gI_mod}\\
    \mathcal{G}^{II}_{\mathrm{mod}} &= \sum_{a,b,c,d=1}^{n} \left( \beta_c E^c \frac{\partial \Phi}{\partial E^c} \delta^c_1 \right) \eta^d_a \frac{\partial^2 \Phi}{\partial E^b \partial E^d} \, dE^a \, dE^b \label{eq:gII_mod}\\
    \mathcal{G}^{III}_{\mathrm{mod}} &= \sum_{a,b=1}^{n} \left( \beta_a E^a \frac{\partial \Phi}{\partial E^a} \right) \delta^b_a\frac{\partial^2 \Phi}{\partial E^a \partial E^b} \, dE^a \, dE^b \label{eq:gIII_mod}
\end{align}

In a 2D thermodynamic system like Bardeen AdS BH metric structure can be expressed as
\begin{equation}
\mathcal{G} = \begin{pmatrix}
    m_{ss} & m_{sg} \\
    m_{sg} & m_{gg}
\end{pmatrix}
\end{equation}

Where the metric elements  $m_{ss}$, $m_{sg}$, and $m_{gg}$ will be determined by the types of metric under consideration (Equation \ref{eq:gI}-\ref{eq:gIII_mod}). In this work, I will analyze the geodesic equation and make a comparative study of both conventional and modified GTD metrics defined by all three metrics.

\section{\label{sec:geo_eq}Thermodynamic Geodesics}

Geodesics represent extremal curves in a Riemannian manifold and play a central role in characterizing the geometric structure of Riemannian space-time. For a manifold endowed with a metric $g_{\mu\nu}$, geodesics are obtained by extremizing the line element
\begin{equation}
\int d\lambda\, \sqrt{g_{\mu\nu}\dot{x}^{\mu}\dot{x}^{\nu}},
\end{equation}
where $\lambda$ is an affine parameter and dots denote derivatives with respect to $\lambda$. Variation of this functional yields the geodesic equation
\begin{equation}
\ddot{x}^{\mu} + \Gamma^{\mu}_{\nu\rho}\dot{x}^{\nu}\dot{x}^{\rho} = 0 ,
\label{eq:geodesic_eq}
\end{equation}
with $\Gamma^{\mu}_{\nu\rho}$ denoting the Christoffel symbols associated with $g_{\mu\nu}$.

Equivalently, the geodesic equations may be derived from the Lagrangian
\begin{equation}
\mathscr{L} = g_{\mu\nu}\dot{x}^{\mu}\dot{x}^{\nu},
\end{equation}
by applying the Euler–Lagrange equations,
\begin{equation}
\frac{d}{d\lambda}\left(\frac{\partial \mathscr{L}}{\partial \dot{x}^{\mu}}\right)
- \frac{\partial \mathscr{L}}{\partial x^{\mu}} = 0 .
\label{eq:geodesic_L}
\end{equation}
Here $\mu = 1,2,\ldots,d$, where $d$ denotes the dimension of the thermodynamic phase space. In the following sections, these equations are employed to analyze the behavior of thermodynamic geodesics defined by conventional and modified GTD metrics.

\section{\label{sec:geo_in_can}Geodesics of Bardeen A\MakeLowercase{d}S BH in canonical ensemble}

\subsection{$\mathcal{G}^I$ vs. $\mathcal{G}^I_{\mathrm{mod}}$ metric}
In conventional GTD metric, structure of  $\mathcal{G}^I$ metric is defined as  $m_{ss} = (2st+ g\phi) \frac{\partial^2 m}{\partial s^2} = \frac{(3 s+1) \left(g^2+s\right) \left(8 g^4+4 g^2 s+s^2 (3 s-1)\right)}{16 s^4}$, $m_{sg} = (2st+ g\phi) \frac{\partial^2 m}{\partial s \partial g} = -\frac{3 g (3 s+1) \left(g^2+s\right) \left(2 g^2-s^2+s\right)}{8 s^3}$, and $m_{gg} = (2st+ g\phi) \frac{\partial^2 m}{\partial g ^2} = \frac{3 (s+1) (3 s+1) \left(g^2+s\right) \left(2 g^2+s\right)}{4 s^2}$. Geodesic equation defined by this metric structure is given by

\begin{multline}
s' \Bigg(\frac{6 g^5 g'}{s^4}+\frac{6 \left(3 g^2+1\right) g^3 g'}{s^3}+\frac{3 g g'}{s}+\frac{3 \left(24 g^2+1\right) g g'}{4 s^2}+ \\
\frac{9}{4} g g'\Bigg) 
+\frac{2 g^6 g'^2}{s^5}+\frac{9 \left(2 g^2+1\right) g^4 g'^2}{4 s^4}+ \frac{9}{4} s \left(g g''+g'^2\right)\\ + 
\frac{-\frac{1}{4} 9 \left(2 g^2+1\right) g^3 g''-\frac{1}{16} \left(360 g^4+96 g^2+1\right) g'^2}{s^2}- \\ 
\frac{3 \left(g \left(8 g^2+1\right) g''+\left(24 g^2+1\right) g'^2\right)}{4 s}- \\
\frac{3 \left(4 g^5 g''+8 g^4 g'^2-g^2 g'^2\right)}{8 s^3}+ \\
\frac{3}{16} \Bigg(4 g \left(3 g^2-2\right) g'' +
\left(36 g^2-11\right) g'^2\Bigg)+ \\
\Bigg(\frac{g^6}{s^4}+
\frac{\frac{3 g^2}{2}- 
\frac{1}{8}}{s}+ \frac{3 \left(12 g^4+g^2\right)}{8 s^2}+ 
\frac{3 \left(2 g^6+g^4\right)}{2 s^3}+ \\
\frac{9 g^2}{8}+\frac{9 s}{8}\Bigg) s'' +
\Bigg(-\frac{2 g^6}{s^5}+\frac{\frac{1}{16}-\frac{3 g^2}{4}}{s^2}- \frac{3 \left(12 g^4+g^2\right)}{8 s^3}- \\
\frac{9 \left(2 g^6+g^4\right)}{4 s^4}+\frac{9}{16}\Bigg) s'^2 = 0
\end{multline}
and
\begin{multline}
\Bigg(\frac{g^6 (3 s+1)}{s^4}+\frac{g^4 (9 s+3)}{2 s^3}+\frac{3 g^2 (s+1) (3 s+1)}{8 s^2}+\frac{9 s}{8}-\frac{1}{8 s}\Bigg) g'' \\
+g' \Bigg(-\frac{g^6 (9 s+4) s'}{s^5}-\frac{9 g^4 (2 s+1) s'}{2 s^4}-\frac{3 g^2 (2 s+1) s'}{4 s^3}+\\
\frac{\left(9 s^2+1\right) s'}{8 s^2}\Bigg) 
+\Big(\frac{g^5 (9 s+3)}{s^4}+\frac{g^3 (9 s+3)}{s^3}+\\
\frac{3 g (s+1) (3 s+1)}{8 s^2}\Big) g'^2 \\
+\frac{3 g^5 \left(s'^2-s (3 s+1) s''\right)}{2 s^4}+ \\
\frac{3 g^3 \left(s \left(3 s^2-8 s-3\right) s''+(2-4 s) s'^2\right)}{4 s^3} \\
+\frac{3 g (s-1) \left(2 s (3 s+1) s''+(3 s-1) s'^2\right)}{8 s^2} = 0
\end{multline}

% I solve geodesic equation numerically and plotted in thermodynamics $s-g$ plane. Behavior of the thermodynamics geodesic is shown in Fig. \ref{fig:g1_g1mod_can}. \\

\begin{figure}
    \centering
    \includegraphics[width=\linewidth]{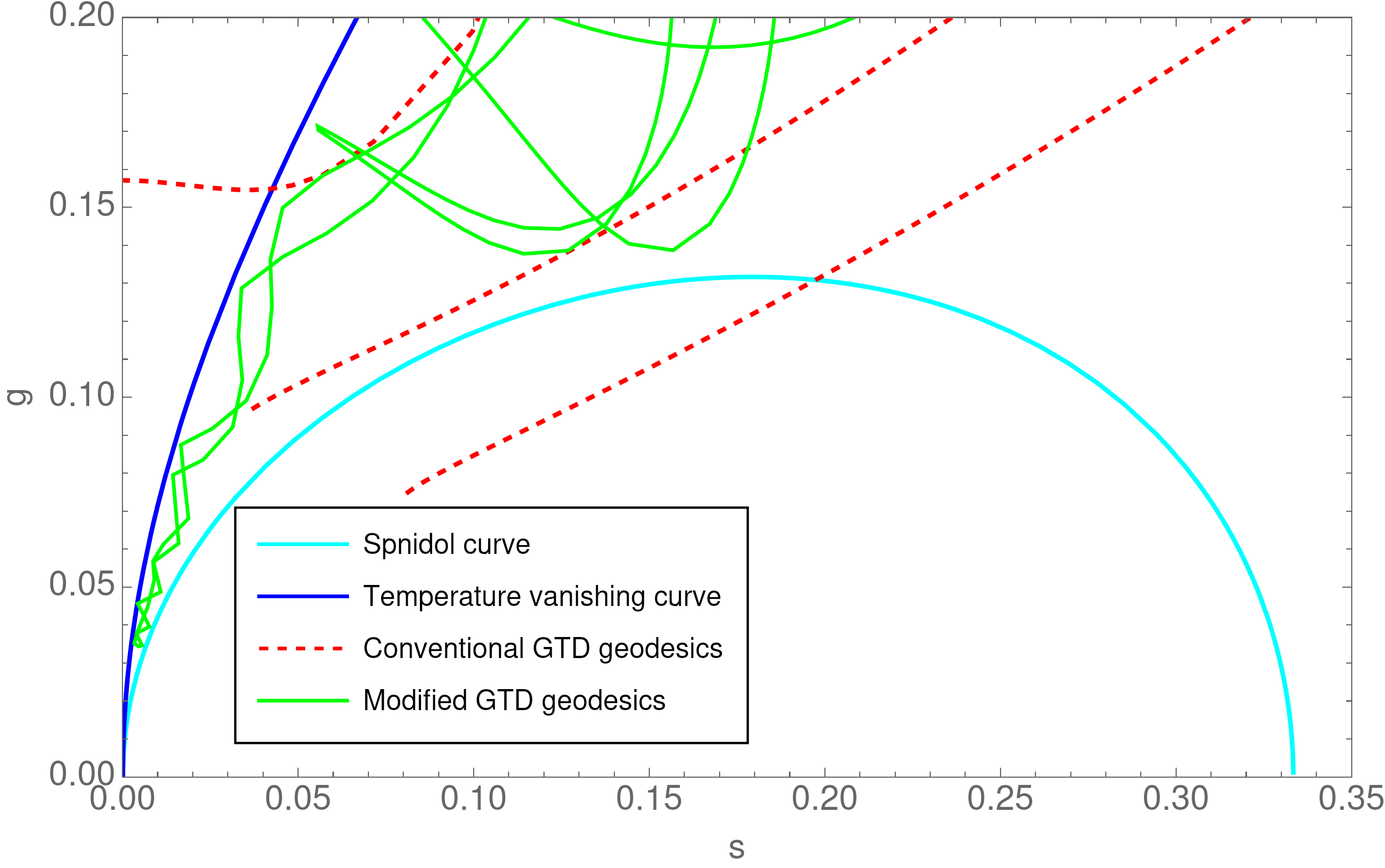}
    \caption{Thermodynamic geodesics of the Bardeen AdS black hole in the canonical ensemble, defined by the $\mathcal{G}^{I}$ and $\mathcal{G}^{I}_{\mathrm{mod}}$ metrics. 
    The $\mathcal{G}^{I}$ geodesic is shown by the red dashed line, while the $\mathcal{G}^{I}_{\mathrm{mod}}$ geodesic is shown by the green solid line. 
    The spinodal and temperature-vanishing curves are indicated by the cyan and blue lines, respectively.}
    \label{fig:g1_g1mod_can}

\end{figure}

Metric element in the modified metric $\mathcal{G}^I_{\mathrm{mod}}$ is given by $m_{ss} = 2st \; \frac{\partial^2 m}{\partial s^2} = \frac{\left(-2 g^2+3 s^2+s\right) \left(8 g^4+4 g^2 s+s^2 (3 s-1)\right)}{16 s^4}$, $m_{sg} = 0$, and $m_{gg} = 2st \; \frac{\partial^2 m}{\partial g^2} = \frac{3 (s+1) \left(2 g^2+s\right) \left(-2 g^2+3 s^2+s\right)}{4 s^2}$. With this metric structure geodesic equation becomes
\begin{multline}
s' \left(-\frac{12 g^5 g'}{s^4}+\frac{3 g g'}{2 s}+\frac{3 \left(8 g^2+1\right) g g'}{2 s^2}\right)-\frac{4 g^6 g'^2}{s^5}+ \\
\frac{3 \left(4 g^2+1\right) g^2 g'^2}{4 s^3}+\frac{\left(6 g^2-1\right) g'^2}{16 s^2}-\frac{9}{16} g'^2+ \\
\left(-\frac{2 g^6}{s^4}+\frac{6 g^2-1}{8 s}+\frac{3 \left(4 g^4+g^2\right)}{4 s^2}+\frac{9 s}{8}\right) s''+ \\
\left(\frac{4 g^6}{s^5}+\frac{1-6 g^2}{16 s^2}-\frac{3 \left(4 g^4+g^2\right)}{4 s^3}+\frac{9}{16}\right) s'^2 = 0
\end{multline}

and 

\begin{multline}
\left(-\frac{2 g^6}{s^4}+\frac{3 g^4}{s^2}+\frac{3 g^2 (s+1)}{4 s^2}+\frac{9 s}{8}-\frac{1}{8 s}\right) g''+ \\
g' \left(\frac{8 g^6 s'}{s^5}-\frac{6 g^4 s'}{s^3}-\frac{3 g^2 (s+2) s'}{4 s^3}+\frac{\left(9 s^2+1\right) s'}{8 s^2}\right)+ \\
\left(-\frac{6 g^5}{s^4}+\frac{6 g^3}{s^2}+\frac{3 g (s+1)}{4 s^2}\right) g'^2+\frac{6 g^5 s'^2}{s^4} \\
-\frac{6 g^3 s'^2}{s^2}-\frac{3 g (s+1) s'^2}{4 s^2} = 0
\end{multline}
I solve geodesic equation numerically and plotted in thermodynamics $s-g$ plane. Behavior of the thermodynamics geodesic is shown in Fig. \ref{fig:g1_g1mod_can}.

\subsection{$\mathcal{G}^{II}$ vs. $\mathcal{G}^{II}_{\mathrm{mod}}$ metric}

Metric structure of conventional $\mathcal{G}^{II}$ is given by $m_{ss} = -(2st +g \phi)  \; \frac{\partial^2 m}{\partial s^2} =-\frac{(3 s+1) \left(g^2+s\right) \left(8 g^4+4 g^2 s+s^2 (3 s-1)\right)}{16 s^4}$, $m_{sg} = 0$, and $m_{gg} = (2st + g\phi) \; \frac{\partial^2 m}{\partial g^2} = \frac{3 (s+1) (3 s+1) \left(g^2+s\right) \left(2 g^2+s\right)}{4 s^2}$. With this metric structure geodesic equation becomes,

\begin{multline}
s' \Bigg(-\frac{6 g^5 g'}{s^4}-\frac{6 \left(3 g^2+1\right) g^3 g'}{s^3}-\frac{3 g g'}{s}- \\
\frac{3 \left(24 g^2+1\right) g g'}{4 s^2}-
\frac{9}{4} g g'\Bigg)-\frac{2 g^6 g'^2}{s^5}-\frac{9 \left(2 g^2+1\right) g^4 g'^2}{4 s^4} \\
-\frac{3 \left(12 g^2+1\right) g^2 g'^2}{8 s^3}+\frac{\left(1-12 g^2\right) g'^2}{16 s^2}+\frac{9}{16} g'^2+ \\
\Bigg(-\frac{g^6}{s^4}+\frac{\frac{1}{8}-\frac{3 g^2}{2}}{s}-\frac{3 \left(12 g^4+g^2\right)}{8 s^2}-\\
\frac{3 \left(2 g^6+g^4\right)}{2 s^3}-\frac{9 g^2}{8}-\frac{9 s}{8}\Bigg) s''+ \\
\Bigg(\frac{2 g^6}{s^5}+\frac{\frac{3 g^2}{4}-\frac{1}{16}}{s^2}+\frac{3 \left(12 g^4+g^2\right)}{8 s^3}+\\
\frac{9 \left(2 g^6+g^4\right)}{4 s^4}-\frac{9}{16}\Bigg) s'^2 = 0
\end{multline}

and 

\begin{multline}
\Bigg(-\frac{g^6 (3 s+1)}{s^4}-\frac{g^4 (9 s+3)}{2 s^3}-\frac{3 g^2 (s+1) (3 s+1)}{8 s^2}-\\
\frac{9 s}{8}+\frac{1}{8 s}\Bigg) g'' \\
+g' \Bigg(\frac{g^6 (9 s+4) s'}{s^5}+\frac{9 g^4 (2 s+1) s'}{2 s^4}+\\
\frac{3 g^2 (2 s+1) s'}{4 s^3}-\frac{\left(9 s^2+1\right) s'}{8 s^2}\Bigg)+\\
\left(-\frac{g^5 (9 s+3)}{s^4}-\frac{g^3 (9 s+3)}{s^3}-\frac{3 g (s+1) (3 s+1)}{8 s^2}\right) g'^2+ \\
\frac{3 g^5 (3 s+1) s'^2}{s^4}+\frac{3 g^3 (3 s+1) s'^2}{s^3}+\frac{3 g (s+1) (3 s+1) s'^2}{8 s^2} = 0
\end{multline}

Metric element in $\mathcal{G}^{II}_{\mathrm{mod}}$ is defined as $m_{ss} = - 2st \; \frac{\partial^2 m}{\partial s^2} = -\frac{\left(-2 g^2+3 s^2+s\right) \left(8 g^4+4 g^2 s+s^2 (3 s-1)\right)}{16 s^4}$, $m_{sg} = 0$, $m_{gg} = 2st \; \frac{\partial^2 m}{\partial g^2} = \frac{3 (s+1) \left(2 g^2+s\right) \left(-2 g^2+3 s^2+s\right)}{4 s^2}$. In the thermodynamics space defined by $\mathcal{G}^{II}_{\mathrm{mod}}$ metric, geodesic equation becomes,

\begin{multline}
s' \left(\frac{12 g^5 g'}{s^4}-\frac{3 g g'}{2 s}-\frac{3 \left(8 g^2+1\right) g g'}{2 s^2}\right)+ \\
\frac{4 g^6 g'^2}{s^5}-\frac{3 \left(4 g^2+1\right) g^2 g'^2}{4 s^3}+\frac{\left(1-6 g^2\right) g'^2}{16 s^2}+\frac{9}{16} g'^2+ \\
\left(\frac{2 g^6}{s^4}+\frac{1-6 g^2}{8 s}-\frac{3 \left(4 g^4+g^2\right)}{4 s^2}-\frac{9 s}{8}\right) s''+ \\
\left(-\frac{4 g^6}{s^5}+\frac{6 g^2-1}{16 s^2}+\frac{3 \left(4 g^4+g^2\right)}{4 s^3}-\frac{9}{16}\right) s'^2 = 0
\end{multline}

and

\begin{multline}
\left(\frac{2 g^6}{s^4}-\frac{3 g^4}{s^2}-\frac{3 g^2 (s+1)}{4 s^2}-\frac{9 s}{8}+\frac{1}{8 s}\right) g''+ \\
g' \left(-\frac{8 g^6 s'}{s^5}+\frac{6 g^4 s'}{s^3}+\frac{3 g^2 (s+2) s'}{4 s^3}-\frac{\left(9 s^2+1\right) s'}{8 s^2}\right)+ \\
\left(\frac{6 g^5}{s^4}-\frac{6 g^3}{s^2}-\frac{3 g (s+1)}{4 s^2}\right) g'^2-\frac{6 g^5 s'^2}{s^4}+\frac{6 g^3 s'^2}{s^2}+ \\
\frac{3 g (s+1) s'^2}{4 s^2} = 0
\end{multline}

\begin{figure}
    \centering
    \includegraphics[width=\linewidth]{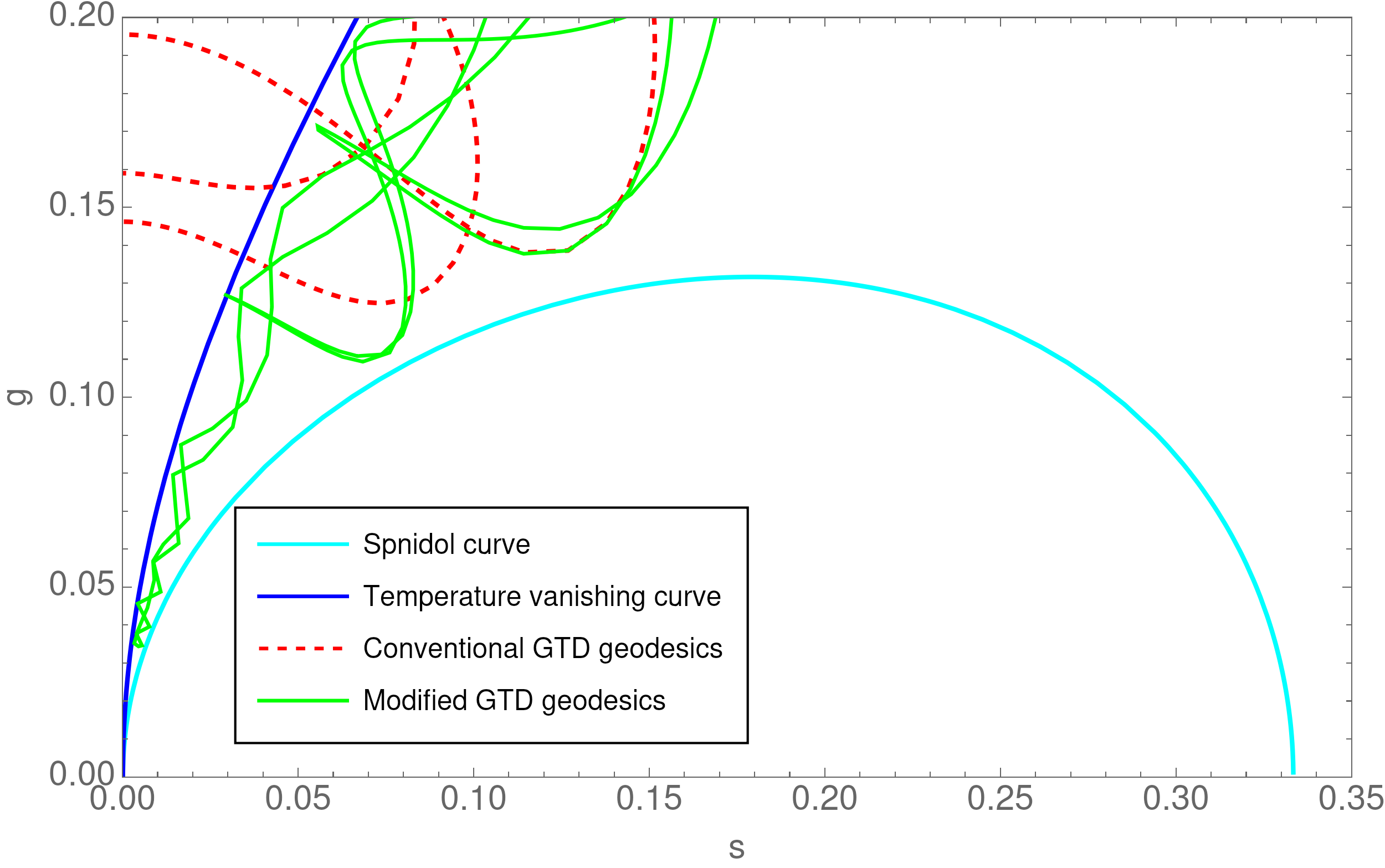}
    \caption{Thermodynamic geodesics of the Bardeen AdS black hole in the canonical ensemble, defined by the $\mathcal{G}^{II}$ and $\mathcal{G}^{II}_{\mathrm{mod}}$ metrics. 
    The $\mathcal{G}^{II}$ geodesic is shown by the red dashed line, while the $\mathcal{G}^{II}_{\mathrm{mod}}$ geodesic is shown by the green solid line. 
    The spinodal and temperature-vanishing curves are indicated by the cyan and blue lines, respectively.}
    \label{fig:g2_g2mod_can}
\end{figure}

The geodesic equations are numerically solved and plotted in $s-g$ plane. Behavior of geodesics is shown in Fig. \ref{fig:g2_g2mod_can}.

\subsection{$\mathcal{G}^{III}$ vs. $\mathcal{G}^{III}_{\mathrm{mod}}$ metric}

Metric in $\mathcal{G}^{III}$ metric is defined by $m_{ss} = 2st \; \frac{\partial^2 m}{\partial s^2} =\frac{\left(-2 g^2+3 s^2+s\right) \left(8 g^4+4 g^2 s+s^2 (3 s-1)\right)}{16 s^4}$, $m_{sg} = \frac{1}{2}(2st + g\phi)\; \frac{\partial^2 m}{\partial s \partial g} =-\frac{3 g (3 s+1) \left(g^2+s\right) \left(2 g^2-s^2+s\right)}{16 s^3}$, and $m_{gg} = g\phi \; \frac{\partial^2 m}{\partial g^2} =\frac{9 g^2 (s+1)^2 \left(2 g^2+s\right)}{4 s^2}$. Geodesic equation defined by this metric structure is given by
\begin{multline}
s' \left(-\frac{12 g^5 g'}{s^4}+\frac{3 g g'}{2 s}+\frac{3 \left(8 g^2+1\right) g g'}{2 s^2}\right) \\
-\frac{4 g^6 g'^2}{s^5}+\frac{9}{8} s \left(g g''+g'^2\right)+ \\
\frac{-\frac{1}{8} 9 \left(2 g^2+1\right) g^3 g''-\frac{1}{16} \left(180 g^4+48 g^2+1\right) g'^2}{s^2}- \\
\frac{3 \left(g \left(8 g^2+1\right) g''+\left(24 g^2+1\right) g'^2\right)}{8 s}- \\
\frac{3 \left(g^5 g''+g^4 g'^2-g^2 g'^2\right)}{4 s^3}+ \\
\frac{3}{16} \left(2 g \left(3 g^2-2\right) g''+ 
\left(18 g^2-7\right) g'^2\right)+ \\
\left(-\frac{2 g^6}{s^4}+\frac{6 g^2-1}{8 s}+\frac{3 \left(4 g^4+g^2\right)}{4 s^2}+\frac{9 s}{8}\right) s'' 
+ \\
\left(\frac{4 g^6}{s^5}+\frac{1-6 g^2}{16 s^2}-\frac{3 \left(4 g^4+g^2\right)}{4 s^3}+\frac{9}{16}\right) s'^2 
= 0
\end{multline}

and 

\begin{multline}
\left(-\frac{2 g^6}{s^4}+\frac{3 g^4}{s^2}+\frac{3 g^2 (s+1)}{4 s^2}+\frac{9 s}{8}-\frac{1}{8 s}\right) g''+ \\
g' \left(\frac{8 g^6 s'}{s^5}-\frac{6 g^4 s'}{s^3}-\frac{3 g^2 (s+2) s'}{4 s^3}+\frac{\left(9 s^2+1\right) s'}{8 s^2}\right)+ \\
\left(-\frac{6 g^5}{s^4}+\frac{6 g^3}{s^2}+\frac{3 g (s+1)}{4 s^2}\right) g'^2+ \\
\frac{g^5 \left(3 (6 s+11) s'^2-3 s (3 s+1) s''\right)}{4 s^4}+ \\
\frac{3 g^3 \left(s \left(3 s^2-8 s-3\right) s''+(6-8 s) s'^2\right)}{8 s^3}+ \\
\frac{3 g \left(3 s^2-2 s-1\right) \left(s s''+s'^2\right)}{8 s^2} =0
\end{multline}

Metric structure of modified $\mathcal{G}^{III}_{\mathrm{mod}}$ metric is same as conventional $\mathcal{G}^{III}$ metric apart from the fact that off diagonal elements in  $\mathcal{G}^{III}_{\mathrm{mod}}$ metric are zero, i.e. $m_{sg} = 0$. With this, geodesic equation will be modified as
\begin{multline}
s' \left(-\frac{12 g^5 g'}{s^4}+\frac{3 g g'}{2 s}+\frac{3 \left(8 g^2+1\right) g g'}{2 s^2}\right) \\
-\frac{4 g^6 g'^2}{s^5}+\frac{3 \left(4 g^2+1\right) g^2 g'^2}{4 s^3}+\frac{\left(6 g^2-1\right) g'^2}{16 s^2}-\frac{9}{16} g'^2+ \\
\left(-\frac{2 g^6}{s^4}+\frac{6 g^2-1}{8 s}+\frac{3 \left(4 g^4+g^2\right)}{4 s^2}+\frac{9 s}{8}\right) s''+ \\
\left(\frac{4 g^6}{s^5}+\frac{1-6 g^2}{16 s^2}-\frac{3 \left(4 g^4+g^2\right)}{4 s^3}+\frac{9}{16}\right) s'^2 = 0
\end{multline}

and 

\begin{multline}
\left(-\frac{2 g^6}{s^4}+\frac{3 g^4}{s^2}+\frac{3 g^2 (s+1)}{4 s^2}+\frac{9 s}{8}-\frac{1}{8 s}\right) g''+ \\
g' \left(\frac{8 g^6 s'}{s^5}-\frac{6 g^4 s'}{s^3}-\frac{3 g^2 (s+2) s'}{4 s^3}+\frac{\left(9 s^2+1\right) s'}{8 s^2}\right)+ \\
\left(-\frac{6 g^5}{s^4}+\frac{6 g^3}{s^2}+\frac{3 g (s+1)}{4 s^2}\right) g'^2+\frac{6 g^5 s'^2}{s^4}-\frac{6 g^3 s'^2}{s^2} \\
-\frac{3 g (s+1) s'^2}{4 s^2} = 0
\end{multline}

\begin{figure}
    \centering
    \includegraphics[width=\linewidth]{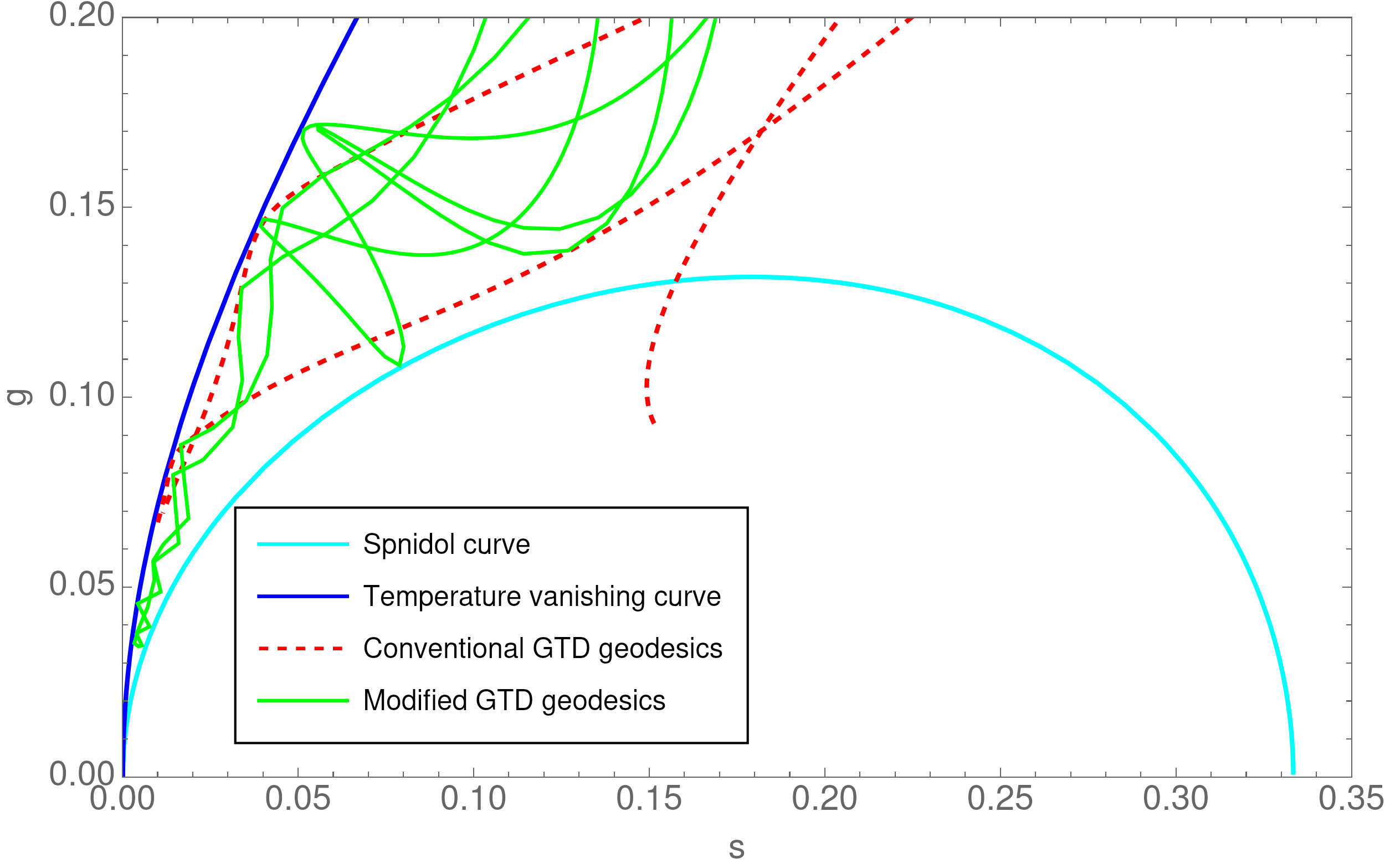}
    \caption{Thermodynamic geodesics of the Bardeen AdS black hole in the canonical ensemble, defined by the $\mathcal{G}^{III}$ and $\mathcal{G}^{III}_{\mathrm{mod}}$ metrics. 
    The $\mathcal{G}^{III}$ geodesic is shown by the red dashed line, while the $\mathcal{G}^{III}_{\mathrm{mod}}$ geodesic is shown by the green solid line. 
    The spinodal and temperature-vanishing curves are indicated by the cyan and blue lines, respectively.}
    \label{fig:g3_g3mod_can}
\end{figure}

The geodesic equation is solved and plotted in $s-g$ plane. Behaviour of geodesic is shown in Fig. \ref{fig:g3_g3mod_can}.

\section{\label{sec:GC_ens}Bardeen A\MakeLowercase{d}S black hole in Grand Canonical Ensembles}
Grand canonical ensemble is the one which allows the fluctuation of all extensive variables. In Bardeen AdS BH, there is two extensive variable entropy $s$, and magnetic charge $g$. In the previous discussion, fluctuations in $s$ were considered while the magnetic charge was held fixed, and the specific heat was therefore defined at constant $g$. This corresponds to a canonical description of the system.
In contrast, the grand canonical ensemble permits fluctuations in the magnetic charge as well. To describe the thermodynamics in this ensemble, it is necessary to introduce a new thermodynamic potential obtained via a Legendre transformation of the canonical mass $m$,
\begin{equation}
M = m - g\phi 
\label{eq:M_and_m}
\end{equation}
where $\phi$ denotes the magnetic potential. A systematic classification of admissible GTD potentials can be found in \cite{Quevedo_2008a}.
Since the magnetic charge is no longer fixed in the grand canonical ensemble, the specific heat cannot be defined at constant $g$. Instead, the appropriate quantity is the specific heat at constant magnetic potential $\phi$. Accordingly, it is convenient to express the Smarr relation as a function of $(s,\phi)$ rather than $(s,g)$, in contrast to the canonical formulation. From equation (\ref{eq:phi}) and (\ref{eq:M_and_m}), potential $M$ can be express as 
\begin{multline}
M = \\
-\frac{1}{6 \sqrt{6} s (s+1)}\Big(-6 s^3+\sqrt{\left(s^2+s\right)^2 \left(9 s^2+18 s+16 \phi ^2+9\right)} \\
-12 s^2-6 s\Big) \\
\sqrt{\frac{3 s^3+\sqrt{\left(s^2+s\right)^2 \left(9 s^2+18 s+16 \phi ^2+9\right)}+6 s^2+3 s}{(s+1)^2}}
\end{multline}
Entropy derivative of this potential will behave as hawking temperature in this phase-space. So temperature is defined as 
\begin{multline}
T = \frac{\partial M}{\partial s} = \frac{1}{12 s (s+1)^2 \sqrt{\frac{6 \sqrt{s^2 (s+1)^2 \left(9 (s+1)^2+16 \phi ^2\right)}}{(s+1)^2}+18 s}} \\
3 \sqrt{s^2 (s+1)^2 \left(9 (s+1)^2+16 \phi ^2\right)}+ \\
s \Big(9 \sqrt{s^2 (s+1)^2 \left(9 (s+1)^2+16 \phi ^2\right)}+ \\
9 s (s (3 s+7)+5)-16 \phi ^2+9\Big)
\end{multline}
In a similar way, specific heat at constant magnetic potential can be defined as 
\begin{multline}
c_\phi = \frac{\left(\frac{\partial M}{\partial s}\right)_\phi}{\left(\frac{\partial^2 M }{\partial s^2}\right)_\phi} = \\
\frac{1}{A} \Bigg(2 s (s+1) \Big(27 \sqrt{s^2 (s+1)^2 \left(9 (s+1)^2+16 \phi ^2\right)}+ \\
s \big(243 s^5+1053 s^4+54 s^3 \left(8 \phi ^2+33\right)+ \\
9 s^2 \left(9 \left(\sqrt{s^2 (s+1)^2 \left(9 (s+1)^2+16 \phi ^2\right)}+18\right)+104 \phi ^2\right)+ \\ 
9 s \left(21 \left(\sqrt{s^2 (s+1)^2 \left(9 (s+1)^2+16 \phi ^2\right)}+3\right)+64 \phi ^2\right)+ \\
72 \phi ^2 \left(\sqrt{s^2 (s+1)^2 \left(9 (s+1)^2+16 \phi ^2\right)}+1\right)+ \\
27 \left(5 \sqrt{s^2 (s+1)^2 \left(9 (s+1)^2+16 \phi ^2\right)}+3\right)-128 \phi ^4\big)\Big)\Bigg)
\end{multline}

Where 
\begin{multline}
    A = s \Bigg(243 s^6+1134 s^5+27 s^4 \left(8 \phi ^2+75\right)+ \\
    27 s^2 \left(8 \sqrt{s^2 (s+1)^2 \left(9 (s+1)^2+16 \phi ^2\right)}+48 \phi ^2+15\right)+ \\
    8 \phi ^2 \left(18 \sqrt{s^2 (s+1)^2 \left(9 (s+1)^2+16 \phi ^2\right)}+16 \phi ^2-9\right)+ 2 s \\
    \left(81 \left(\sqrt{s^2 (s+1)^2 \left(9 (s+1)^2+16 \phi ^2\right)}-1\right)+256 \phi ^4+216 \phi ^2\right) \\
    +9 s^3 \left(9 \left(\sqrt{s^2 (s+1)^2 \left(9 (s+1)^2+16 \phi ^2\right)}+20\right)+112 \phi ^2\right) \\
    -81\Bigg) -27 \sqrt{s^2 (s+1)^2 \left(9 (s+1)^2+16 \phi ^2\right)}
\end{multline}
In grand canonical ensemble also, spinodal curve and temperature vanishing curve can be obtained by solving $T=0$, and $c_{\phi} \rightarrow \infty $ respectively. This gives spinodal curve as
\begin{multline}
\phi = \Bigg(-\frac{27 s^2}{2 \left(16 s^2+8 s+1\right)}-\frac{9 s}{8 \left(16 s^2+8 s+1\right)}+ \\
\frac{9}{16 \left(16 s^2+8 s+1\right)} - 
\frac{27 s^5}{4 \left(16 s^2+8 s+1\right)}-\frac{369 s^4}{16 \left(16 s^2+8 s+1\right)}\\
-\frac{225 s^3}{8 \left(16 s^2+8 s+1\right)} 
+\frac{1}{8 \left(16 s^2+8 s+1\right)} \\
\Big(9 (-12 s^9-74 s^8-190 s^7-257 s^6-184 s^5-47 s^4+ \\
26 s^3+25 s^2+8 s+1)^{1/2}\Big)\Bigg)^{1/2}
\end{multline}
Similarly equation of temperature vanishing curve becomes,
\begin{equation}
    \phi = \frac{3}{4} \sqrt{3} (s+1)^{3/2} \sqrt{3 s+1}
\end{equation}

\begin{figure}
    \centering
    \includegraphics[width=\linewidth]{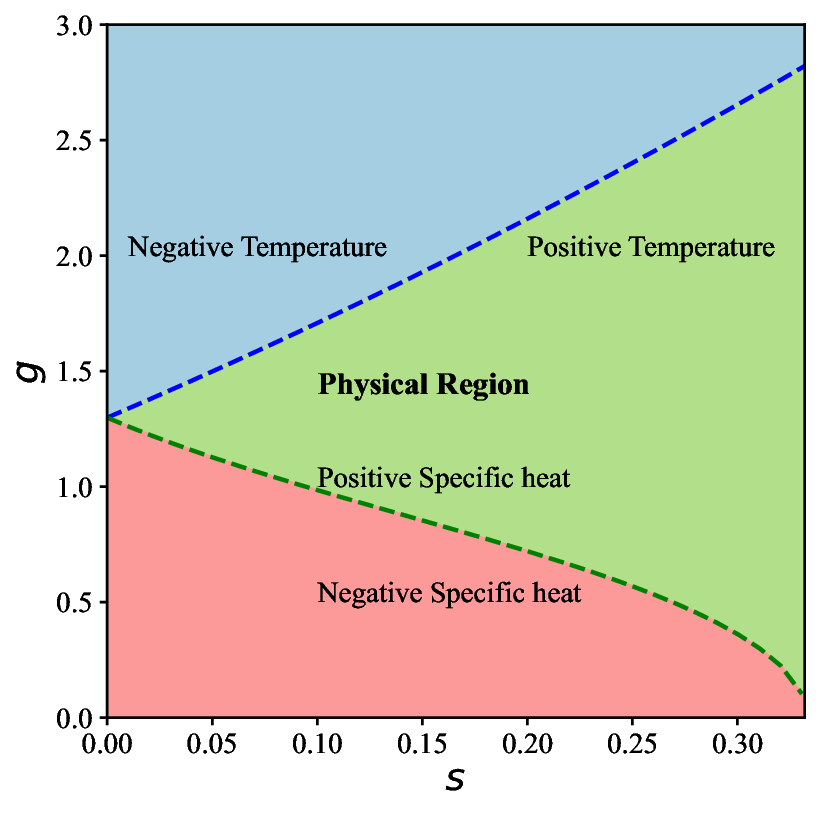}
     \caption{$s$--$\phi$ thermodynamic phase space of the Bardeen AdS black hole. The green dotted curve denotes the spinodal line separating regions of positive and negative specific heat, while the blue dotted curve represents the temperature-vanishing line. The green shaded area corresponds to the physical region with positive temperature and positive specific heat; red and blue shaded regions indicate negative specific heat and negative temperature, respectively.} 
    \label{fig:physical_region_GC}
\end{figure}

Physical region of Bardeen AdS BH defined by the physical boundaries is shown in Fig. \ref{fig:physical_region_GC}. 
Generalized metric structure of Bardeen AdS BH in grand canonical ensemble with potential $M$ can be expressed as,
\begin{equation}
\mathcal{G} = \begin{pmatrix}
    M_{ss} & M_{s\phi} \\
    M_{s\phi} & M_{\phi\phi}
\end{pmatrix}
\end{equation}
In constructing the metric components, it is important to note that in the grand canonical ensemble there is only one extensive variable, namely the entropy $s$, since the magnetic potential $\phi$—conjugate to the magnetic charge—plays the role of an intensive parameter. Consequently, the degrees of homogeneity are $\beta_s = 2$ and $\beta_\phi = 0$.
\section{\label{sec:geo_in_GC}Geodesics of Bardeen A\MakeLowercase{d}S black hole in grand canonical ensemble}

\subsection{$\mathcal{G}^{I}$ vs. $\mathcal{G}^I_{\mathrm{mod}}$ metric}

In grand canonical ensemble, metric elements of Bardeen AdS BH with $\mathcal{G}^I$ metric structure is given by
 $M_{ss}= 2st \; \frac{\partial ^2 M}{\partial s^2} = $
 \begin{multline}
     \frac{1}{B} 
     \Bigg(3 \sqrt{s^2 (s+1)^2 \left(9 (s+1)^2+16 \phi ^2\right)}+ \\
     s \Big(9 \sqrt{s^2 (s+1)^2 \left(9 (s+1)^2+16 \phi ^2\right)}+9 s (s (3 s+7)+5) \\
     -16 \phi ^2+9\Big)\Bigg) \\
     \Bigg(s \Big(243 s^6+1134 s^5+27 s^4 \left(8 \phi ^2+75\right)+ \\
     27 s^2 \left(8 \sqrt{s^2 (s+1)^2 \left(9 (s+1)^2+16 \phi ^2\right)}+48 \phi ^2+15\right)+8 \phi ^2 \\
     \left(18 \sqrt{s^2 (s+1)^2 \left(9 (s+1)^2+16 \phi ^2\right)}+16 \phi ^2-9\right)+ \\
     2 s \Big(81 \left(\sqrt{s^2 (s+1)^2 \left(9 (s+1)^2+16 \phi ^2\right)}-1\right)+256 \phi ^4+ \\
     216 \phi ^2\Big)+9 s^3 \Big(9 \left(\sqrt{s^2 (s+1)^2 \left(9 (s+1)^2+16 \phi ^2\right)}+20\right)+ \\
     112 \phi ^2\Big)-81\Big)-27 \sqrt{s^2 (s+1)^2 \left(9 (s+1)^2+16 \phi ^2\right)}\Bigg) \nonumber
 \end{multline}. 
 \begin{multline}
         M_{s\phi} = 2st \frac{\partial ^2 M}{\partial s \partial \phi} =\\
         \frac{1}{C} s \phi  \Bigg(s \Big(3 \sqrt{s^2 (s+1)^2 \left(9 (s+1)^2+16 \phi ^2\right)}+ \\
         9 s \left(s^2+s-1\right)-16 \phi ^2-9\Big)- \\
         3 \sqrt{s^2 (s+1)^2 \left(9 (s+1)^2+16 \phi ^2\right)}\Bigg) \\
         \Bigg(3 \sqrt{s^2 (s+1)^2 \left(9 (s+1)^2+16 \phi ^2\right)}+s \\
         \Big(9 \sqrt{s^2 (s+1)^2 \left(9 (s+1)^2+16 \phi ^2\right)}+9 s (s (3 s+7)+5) \\
         -16 \phi ^2+9\Big)\Bigg) \nonumber
 \end{multline}
 \begin{multline}
     M_{\phi \phi} =  2st \frac{\partial ^2 M}{\partial \phi^2} = \\\frac{s \left(-\frac{s \left(9 \sqrt{s^2 (s+1)^2 \left(9 (s+1)^2+16 \phi ^2\right)}+9 s (s (3 s+7)+5)-16 \phi ^2+9\right)}{\sqrt{s^2 (s+1)^2 \left(9 (s+1)^2+16 \phi ^2\right)}}-3\right)}{18 (s+1)} \nonumber
 \end{multline}
 
\begin{figure}
    \centering
    \includegraphics[width=\linewidth]{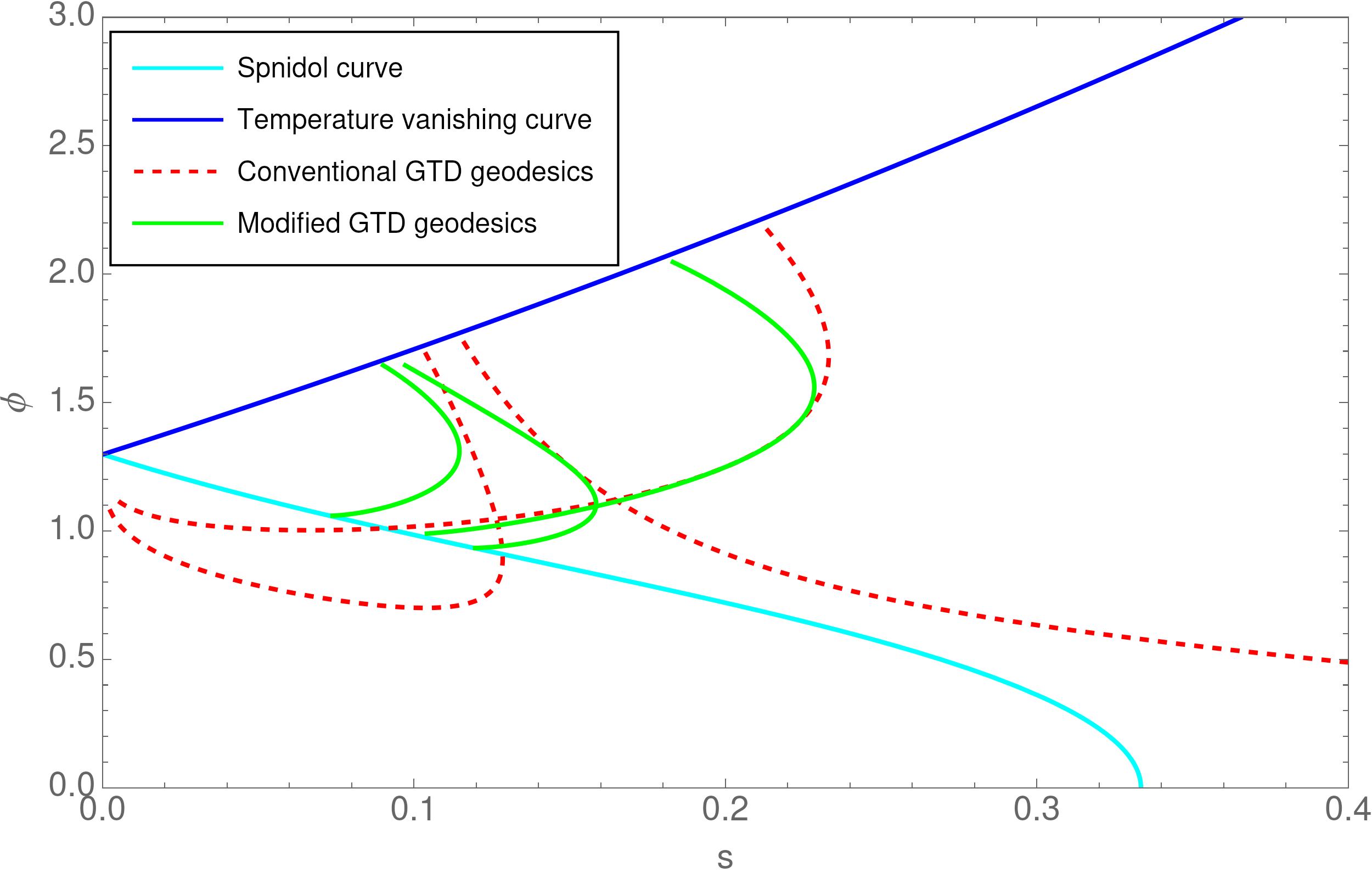}
    \caption{Thermodynamic geodesics of the Bardeen AdS black hole in the grand canonical ensemble, defined by the $\mathcal{G}^{I}$ and $\mathcal{G}^{I}_{\mathrm{mod}}$ metrics. 
    The $\mathcal{G}^{I}$ geodesic is shown by the red dashed line, while the $\mathcal{G}^{I}_{\mathrm{mod}}$ geodesic is shown by the green solid line. 
    The spinodal and temperature-vanishing curves are indicated by the cyan and blue lines, respectively.}
    \label{fig:g1_g1mod_GC}
\end{figure}

 Where
 \begin{multline}
     B = 432 (s+1) \sqrt{s^2 (s+1)^2 \left(9 (s+1)^2+16 \phi ^2\right)} \\
     \left(\sqrt{s^2 (s+1)^2 \left(9 (s+1)^2+16 \phi ^2\right)}+3 s (s+1)^2\right)^2  \nonumber
 \end{multline}.
 \begin{multline}
     C = 18 \sqrt{s^2 (s+1)^2 \left(9 (s+1)^2+16 \phi ^2\right)} \\
     \left(\sqrt{s^2 (s+1)^2 \left(9 (s+1)^2+16 \phi ^2\right)}+3 s (s+1)^2\right)^2 \nonumber
 \end{multline}
 With this metric structure the geodesic equations can be obtained by following equation (\ref{eq:geodesic_eq}) or (\ref{eq:geodesic_L}). I solve these geodesic equations numerically. Behavior of geodesic defined by $\mathcal{G}^I$ metric is shown in Fig. \ref{fig:g1_g1mod_GC}. \\
Metric structure of $\mathcal{G}^I_{\mathrm{mod}}$ metric is same as that of $\mathcal{G}^I$ metric, apart from the fact that ius this case the off diagonal element $M_{s\phi} =0$. With the modified metric structure geodesic equation can be obtained by following equation (\ref{eq:geodesic_eq}) or (\ref{eq:geodesic_L}). I have solved the geodesic equation defined by this metric structure numerically and plotted in $s-g$ plane. Behavior of thermodynamics geodesic defined by $\mathcal{G}^I_{\mathrm{mod}}$ metric is shown in Fig. \ref{fig:g1_g1mod_GC}.

\subsection{$\mathcal{G}^{II}$ and $\mathcal{G}^{II}_{\mathrm{mod}}$ metric}
\begin{figure}
    \centering
    \includegraphics[width=\linewidth]{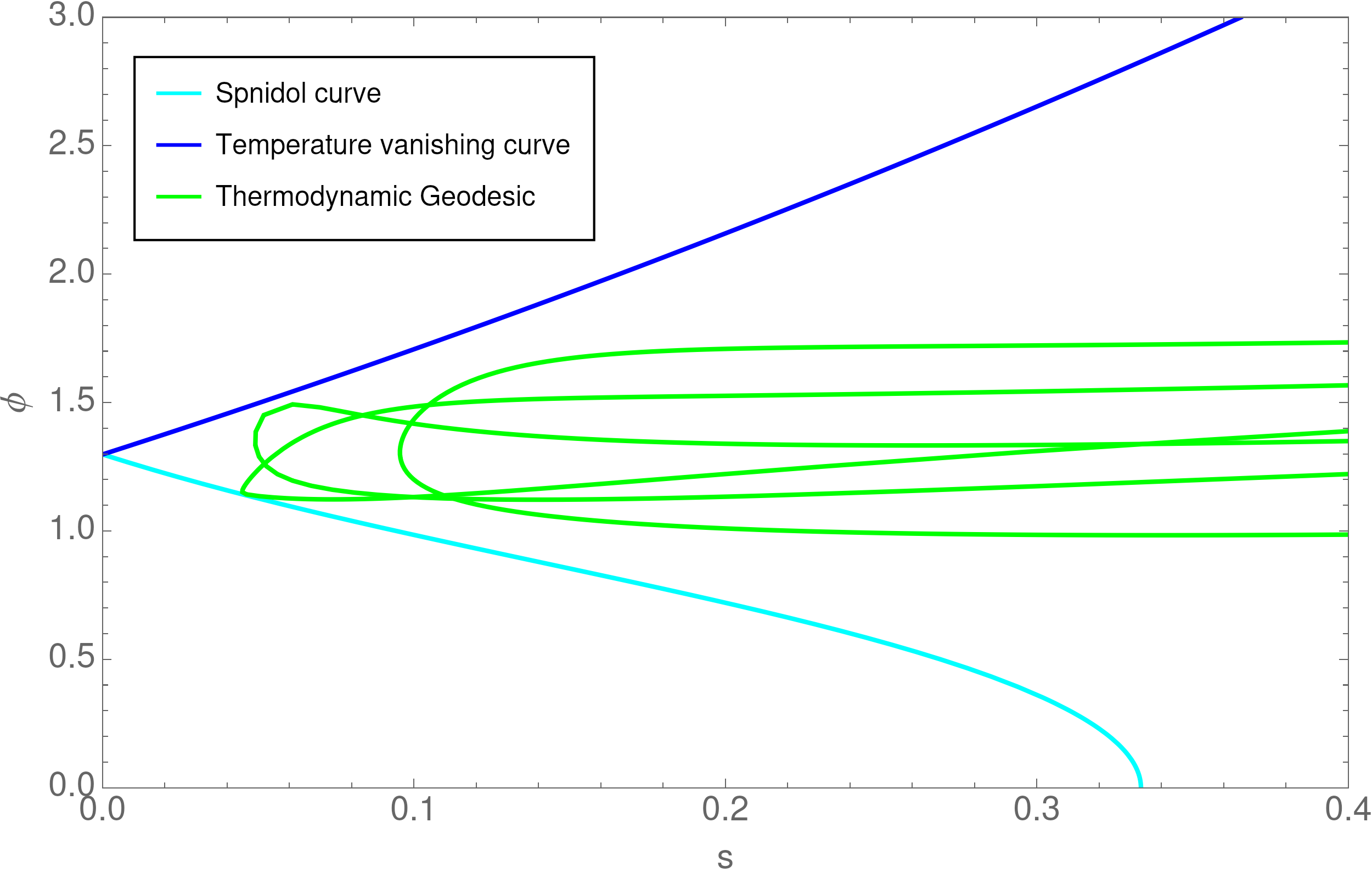}
    \caption{Thermodynamics geodesic of Bardeen AdS black hole in canonical ensemble defined by $\mathcal{G}^{II}$ and $\mathcal{G}^{II}_{\mathrm{mod}}$ metric. The geodesic defined by $\mathcal{G}^{I}$ or $\mathcal{G}^{I}_{\mathrm{mod}}$ metric is shown by green solid line. Spinodal curve and temperature vanishing curve is shown by cyan and blue colored line respectively.}
    \label{fig:g2_g2mod_GC}
\end{figure}

It is important to note that in Grand canonical ensemble, there is only one extensive variable $s$, hence degree of homogeneity for the intensive parameter $\phi$ is zero. This behavior make the $\mathcal{G}^{II}$ metric structure same as $\mathcal{G}^{II}_{\mathrm{mod}}$ metric, although there is structural difference between these two metric. Hence metric element of $\mathcal{G}^{II}$ as well as $\mathcal{G}^{II}_{\mathrm{mod}}$ metric can be expressed as 
$M_{ss}= - 2st \; \frac{\partial ^2 M}{\partial s^2} = $
 \begin{multline}
    - \frac{1}{B} 
     \Bigg(3 \sqrt{s^2 (s+1)^2 \left(9 (s+1)^2+16 \phi ^2\right)}+ \\
     s \Big(9 \sqrt{s^2 (s+1)^2 \left(9 (s+1)^2+16 \phi ^2\right)}+9 s (s (3 s+7)+5) \\
     -16 \phi ^2+9\Big)\Bigg) \\
     \Bigg(s \Big(243 s^6+1134 s^5+27 s^4 \left(8 \phi ^2+75\right)+ \\
     27 s^2 \left(8 \sqrt{s^2 (s+1)^2 \left(9 (s+1)^2+16 \phi ^2\right)}+48 \phi ^2+15\right)+8 \phi ^2 \\
     \left(18 \sqrt{s^2 (s+1)^2 \left(9 (s+1)^2+16 \phi ^2\right)}+16 \phi ^2-9\right)+ \\
     2 s \Big(81 \left(\sqrt{s^2 (s+1)^2 \left(9 (s+1)^2+16 \phi ^2\right)}-1\right)+256 \phi ^4+ \\
     216 \phi ^2\Big)+9 s^3 \Big(9 \left(\sqrt{s^2 (s+1)^2 \left(9 (s+1)^2+16 \phi ^2\right)}+20\right)+ \\
     112 \phi ^2\Big)-81\Big)-27 \sqrt{s^2 (s+1)^2 \left(9 (s+1)^2+16 \phi ^2\right)}\Bigg) \nonumber
 \end{multline}. 
 
 \begin{multline}
     M_{\phi \phi} =  2st \frac{\partial ^2 M}{\partial \phi^2} = \\
     \frac{s \left(-\frac{s \left(9 \sqrt{s^2 (s+1)^2 \left(9 (s+1)^2+16 \phi ^2\right)}+9 s (s (3 s+7)+5)-16 \phi ^2+9\right)}{\sqrt{s^2 (s+1)^2 \left(9 (s+1)^2+16 \phi ^2\right)}}-3\right)}{18 (s+1)} \nonumber
 \end{multline}
 With $B$ having the same meaning as previously defined.
 I solve geodesic equation defined by this metric structure. Behavior of thermodynamics geodesic defined by $\mathcal{G}^{II}$ or $\mathcal{G}^{II}_{\mathrm{mod}}$ metric is shown in Fig. \ref{fig:g2_g2mod_GC}.

\section{\label{sec:Result}Discussion and conclusion}
I analyze the behavior of conventional and modified Geometrothermodynamics (GTD) metrics in AdS spacetime and across different thermodynamic ensembles. The results show that conventional GTD metrics generically fail to confine thermodynamic geodesics within the physical region, revealing an intrinsic limitation of their metric structure. This failure persists irrespective of the spacetime background or ensemble choice, indicating that it is a structural feature of the conventional GTD framework rather than a model-dependent artifact.
For AdS black holes, thermodynamic geodesics associated with conventional GTD metrics are found to cross physical boundaries of the phase space. In particular, geodesics defined by the $\mathcal{G}^I$ metric intersect both the spinodal and temperature-vanishing curves (Fig. \ref{fig:g1_g1mod_can}), those defined by the $\mathcal{G}^{II}$ metric cross the temperature-vanishing curve (Fig. \ref{fig:g2_g2mod_can}), and those associated with the $\mathcal{G}^{III}$ metric cross the spinodal curve (Fig. \ref{fig:g3_g3mod_can}). In contrast, geodesics generated by the modified GTD metrics remain confined within the physical region, exhibiting either turn around behavior or geodesic incompleteness. Figures \ref{fig:g1_g1mod_can}–\ref{fig:g3_g3mod_can} present representative comparisons between geodesics obtained from conventional and modified GTD metrics.
As shown in Paper I, the physical origin of geodesic non-confinement lies in the underlying metric structure. While turn around behavior or geodesic incompleteness is expected near the temperature-vanishing or spinodal curves, a thermodynamically consistent geodesic should display such behavior specifically at points where the thermodynamic curvature diverges. By explicitly evaluating the curvature scalar, I find that the observed confinement of geodesics occurs precisely at these curvature singularities, establishing a direct geometric link between curvature divergence and geodesic confinement.
I further investigate the behavior of GTD metrics in different thermodynamic ensembles. In the grand canonical ensemble, geodesics defined by the modified GTD metrics remain confined within a single thermodynamic phase. While geodesics associated with the conventional $\mathcal{G}^{I}$ metric cross the spinodal curve, those defined by the corresponding modified metric remain entirely within the physical region. For systems characterized by a single extensive variable, the metric structure of the conventional $\mathcal{G}^{II}$ metric coincides with that of the modified $\mathcal{G}^{II}_{\mathrm{mod}}$ metric, and geodesics in both cases remain confined. In this situation, the modified $\mathcal{G}^{III}_{\mathrm{mod}}$ metric cannot be defined.
These results demonstrate that geodesic confinement within the physical region provides a stringent and physically motivated criterion for assessing the adequacy of a thermodynamic metric, a criterion that is consistently satisfied by the modified GTD metrics.

\section*{Acknowledgements}
I sincerely thank the anonymous reviewer for the valuable suggestions, which have significantly improved the quality
of the manuscript.

% \appendix

% \section{Appendix title 1}
% %% \label{}

% \section{Appendix title 2}

\bibliographystyle{elsarticle-harv} 
\bibliography{example}



%% For example:

%% \bibitem[Aladro et al.(2015)]{Aladro15} Aladro, R., Martín, S., Riquelme, D., et al. 2015, \aas, 579, A101

%%\end{thebibliography}

\end{document}